# Intrinsic vs. Extrinsic Magnetic Transitions in $Sr_3Ru_2O_7$ films

Rashmi Choudhary[1,*], Anil Rajapitamahuni[1], Silu Guo[1], Qianni Jiang[2], Jiun-Haw Chu[2], K. Andre Mkhoyan[1], Bharat Jalan[1,*]

[1]Department of Chemical Engineering and Materials Science, University of Minnesota, Minneapolis, Minnesota 55455, USA

[2]Department of Physics, University of Washington, Seattle, Washington 98105, USA

*Corresponding authors: choud140@umn.edu, bjalan@umn.edu

**Abstract**

In scientific research, both positive and negative results play crucial role in advancing the field. Negative results provide valuable insights that can guide future experiments and prevent repeated failures. Here we present our growth attempts of $Sr_3Ru_2O_7$ thin films using the hybrid molecular beam epitaxy. X-ray diffraction suggests nominally phase-pure films. A combination of magnetoresistance and magnetization measurements exhibits an onset of ferromagnetism at 170 K and 100 K, along with a metamagnetic-like transition at 40 K. These results could initially be interpreted as intrinsic behavior of strain-engineered $Sr_3Ru_2O_7$ films. However, detailed microstructural analysis reveals intergrowths of $Sr_2RuO_4$, $Sr_4Ru_3O_{10}$, and $SrRuO_3$ phases, dispersed throughout the film. Our findings suggest that the $Sr_3Ru_2O_7$ films are likely paramagnetic, with the observed ferromagnetism arising from the $Sr_4Ru_3O_{10}$ and $SrRuO_3$ phases. Our results highlight the need for detailed microstructural analysis when interpreting new material properties influenced by strain and microstructure.

**Introduction**

The Ruddlesden-Popper (RP) series $Sr_{n+1}Ru_nO_{3n+1}$ has attracted considerable attention due to its strongly correlated electronic behavior, which gives rise to a wide range of physical phenomena. Notably, the n = 1 member, $Sr_2RuO_4$, exhibits unconventional superconductivity, while the n = 3 and ∞ phases, $Sr_4Ru_3O_{10}$ and $SrRuO_3$, respectively, display ferromagnetic ordering.[1–5] Additionally, $Sr_4Ru_3O_{10}$ is recognized for undergoing a metamagnetic transition below approximately 50 K.[4–6] The n = 2 phase, $Sr_3Ru_2O_7$, has been experimentally shown to be a paramagnetic metal that exhibits an electronic nematic phase at temperatures below 100 mK.[7,8] This nematicity is driven by a metamagnetic transition that occurs near a quantum critical point.[9–11]

In contrast, density functional theory calculations suggest competing magnetic orders in $Sr_3Ru_2O_7$, including a ferromagnetic ground state, challenging the experimental findings.[12] Several other studies report diverse magnetic ground states in $Sr_3Ru_2O_7$, such as paramagnetism, ferromagnetism, and antiferromagnetism.[7,13–18] Notably, an electronic nematic phase transition has been observed in thin films at 15 K, which is significantly higher than the transition temperature observed in bulk, and is attributed to epitaxial strain effects.[13,15] However, it remains unclear whether these magnetic transitions are intrinsic to $Sr_3Ru_2O_7$ or governed by factors such as dimensionality, strain, or the intermixing of other RP ruthenate phases. To resolve these ambiguities, high-quality crystals with fewer defects are crucial. This has spurred considerable interest in synthesizing $Sr_3Ru_2O_7$ thin films, which offer the potential to explore its magnetic and electronic ground states under the effects of reduced dimensionality and epitaxial strain.[13–16]

The similar formation energies of various RP ruthenates make it exceptionally difficult to grow pure $Sr_3Ru_2O_7$ thin films, often resulting in phase intermixing.[19] X-ray diffraction may not adequately resolve these RP stacking faults, as the diffraction peaks for the different phases occur at 2θ positions that are closely spaced. When secondary phases are present in small amounts, the resulting peaks may broaden, making it challenging to distinguish between distinct phases. Similarly, scanning transmission electron microscopy (STEM) images obtained from a few small-areas may fail to detect these phases, as they can easily be missed. The intermixing, however small, can complicate the interpretation of a material's intrinsic electronic and magnetic properties, especially for $Sr_3Ru_2O_7$, where the effect of strain on its properties is not yet well understood.

In this article, we present results from our attempts to grow phase-pure and single-crystalline $Sr_3Ru_2O_7$ thin films, noting multiple magnetic transitions that appear to be attributable to impurity phases. We investigated the electrical and magnetotransport properties of these films across a range of temperatures, identifying three distinct transitions at 170 K, 100 K, and 40 K. Using large-area STEM imaging, we detected intermixing from other RP phases, suggesting that these observed magnetic transitions may not be intrinsic to $Sr_3Ru_2O_7$ and could be explained by the presence of multiple phases.

**Results and Discussion**

A 60-nm-thick $Sr_3Ru_2O_7$ film was grown on an LSAT (001) substrate. $Sr_3Ru_2O_7$ crystallizes in a tetragonal structure with lattice parameters of a = b = 3.89 Å and c = 20.7 Å, as depicted in Figure 1(a).[20] High-resolution X-ray diffraction (HRXRD) characterization of the film, shown in Figure 1(b), reveals a single-crystalline, (001) out-of-plane oriented, and nominally single-phase structure. However, the diffraction peaks appear broadened, suggesting the possibility of RP impurity phases. The intensity of the (008) plane reflection is significantly reduced, while the (006) reflection is absent. Figure 1(c) displays the reflection high-energy electron diffraction (RHEED) patterns of the film along the [100] azimuthal direction before and after growth. The streaky RHEED patterns indicate a smooth surface, high crystalline quality, and an epitaxial relationship with the substrate. The reciprocal space map of the film, shown in Figure 1(d), confirms that the $Sr_3Ru_2O_7$ film is fully coherent on the LSAT substrate. However, the out-of-plane lattice parameter of the film (21.0 Å) is slightly larger than the expected value (20.8 Å) for a fully strained film. This is likely due to the presence of defects such as Ru or oxygen vacancies, intermixing with other RP phases, or a combination of both.[21–23]

Figure 2 presents the electrical and magnetic characterization of the same $Sr_3Ru_2O_7$ film. Figure 2(a) depicts the temperature-dependent in-plane electrical resistivity of the film. The residual resistivity ratio (RRR), defined as $\rho_{300K}/\rho_{2K}$, for our film is 22, which is among the highest reported values for $Sr_3Ru_2O_7$ films. While a higher RRR typically suggests improved crystalline quality and low defect density, caution must be exercised as a small amount of metallic phase-impurity can significantly influence this measurement. The resistivity curve in Figure 2(a) also exhibits weak kink-like features at different temperatures. To better visualize these features, we plot, in Figure 2(b), the derivative of the electrical resistivity with respect to temperature (dρ/dT), revealing three distinct regions marked by vertical dashed lines. The corresponding temperatures suggest the occurrence of possible phase transitions.

To assess whether these transitions have a magnetic origin, we conducted temperature-dependent measurements of the out-of-plane magnetization (M) on the same sample, as shown in Figure 2(c). Figure 2(d) presents the derivative of the magnetization data with respect to temperature (dM/dT) for better visualization of the subtle change in M vs. T. Similar to dρ/dT vs. T, the dM/dT as a function of temperature reveals the same three distinct regions marked by vertical dashed lines. These vertical lines correspond to magnetic phase transitions occurring at 170 K, 100 K, and 40 K. Sections of Figures 2(b) and 2(d) are color-coded to visually differentiate these transitions. It is important to note that the magnetization is reported in units of emu and not converted to Bohr magnetons, as it remains uncertain whether these transitions are intrinsic to the $Sr_3Ru_2O_7$ phase or a result of phase impurities, as discussed later.

For the same film, figure 3 presents the anomalous Hall effect (AHE) resistance ($R_{AHE}$), magnetoresistance (MR), and magnetization as a function of temperature. All three measurements exhibit hysteresis at T ≤ 120 K accompanied by a negative MR at 30 K ≤ T < 120 K. These observations are consistent with an itinerant ferromagnetic state, suggesting that the magnetic transition at 170 K is associated with the onset of ferromagnetism. Additionally, both the magnetization and $R_{AHE}$ begin to exhibit a step-like feature at T ≤ 80 K, suggesting the transition at 100 K is likely related to the onset of a second ferromagnetic phase and the possibility of two magnetic domains with different coercivity. The third transition at ~ 40 K, as depicted in Figure 2, coincides with the onset of an unusual magnetoresistance behavior: at T ≤ 30 K, the MR becomes positive, initially increasing with the magnetic field before decreasing, resulting in a peak. This peak shifts to a higher field with decreasing temperature. In the literature, this MR behavior has been attributed to strain driven electronic nematicity in $Sr_3Ru_2O_7$ films.[13] However, the electronic nematic transition temperature observed in $Sr_3Ru_2O_7$ films is significantly higher than that in bulk single crystals, where it occurs below 100 mK.[8] In bulk single crystals, this transition is reported to originate from a metamagnetic transition giving rise to spin density wave close to a quantum critical point.[11] This spin density wave leads to a sharp increase in magnetoresistance close to a quantum critical field.[8] The critical field ($H_c$) at which a sharp increase in MR is observed is a function of the tilt angle between the ab-plane and applied magnetic field. $H_c$ varies smoothly from 5.1 T for H || ab (defined as 0°) to 7.87 T for H || c (defined as 90°).[8] An evidence of the electronic nematic phase, the in-plane MR along the [100] and [010] crystalline directions becomes anisotropic at $H_c$.[8] The anisotropy in MR along the [100] and [010] directions is only observed when the angle between H and ab-plane is between 0° and 60°, where it is more pronounced for some angles than others.[8]

To investigate whether the unusual MR observed in our film results from metamagnetism or electronic nematicity, we measured the in-plane MR along the [100] and [010] crystalline directions at 1.7 K. As shown in Supplementary Figure S1, the MR along these two directions was measured at different tilt angles between the applied field and the c-axis. From Figure S1, we observe that the peak in MR appears at ~5.2 T for H || ab and shifts smoothly to higher fields as the applied field is tilted away from the ab-plane toward the c-axis, reaching ~7.9 T for H || c. Notably, these fields are very close to the critical fields for metamagnetism in bulk single-crystal $Sr_3Ru_2O_7$. However, we observed little to no anisotropy between the two in-plane directions for 0° < H∠ab < 60°, suggesting isotropic MR behavior and implying the absence of electronic nematicity. We hypothesize that this unusual MR behavior likely originates from the metamagnetism associated with the $Sr_3Ru_2O_7$ phase. However, the absence of nematicity and the specific shape of MR could also be influenced by the presence of other RP phases and the interfaces between them. For instance, a metamagnetic transition below 50 K has also been reported in the $Sr_4Ru_3O_{10}$ phase.[4–6] In $Sr_4Ru_3O_{10}$, at T ≤ 50 K, the magnetic moments, while remaining ferromagnetically aligned out-of-plane, become "locked" into a canted antiferromagnetic arrangement in-plane.[6] A more detailed study is required to investigate the origin of this transition.

Figures 4 presents high-angle annular dark-field (HAADF) STEM images of the representative $Sr_3Ru_2O_7$ film of identical thickness. Figure 4(b) shows a small-area, atomic-resolution image of the film revealing presence of phase-pure $Sr_3Ru_2O_7$, consistent with the result of XRD. However, larger-area STEM images, as shown in Figure 4(a), reveal significant intermixing with other RP phases. Figure 4(c) shows the image in Figure 4(a) with color coding to highlight different phases, including $Sr_2RuO_4$, $Sr_4Ru_3O_{10}$, and $SrRuO_3$, alongside $Sr_3Ru_2O_7$. While these phases exhibit good crystalline quality, the intermixing between them is substantial due to their similar formation energies.[19] Based on our STEM analysis, it is plausible that the observed magnetic transitions at 170 K and 100 K are related to the ferromagnetism arising from the presence of $Sr_4Ru_3O_{10}$ and $SrRuO_3$ phases. Both phases are known to be ferromagnetic, with Curie temperatures of approximately 160 K for $SrRuO_3$ and 105 K for $Sr_4Ru_3O_{10}$.[4,24]

**Summary**

In conclusion, we present our attempts to grow phase-pure $Sr_3Ru_2O_7$ films via hybrid MBE, accompanied by an investigation into their electrical and magnetotransport properties. Our results reveal three distinct magnetic transitions occurring at approximately 170 K, 100 K and 40 K. Detailed STEM imaging uncovered significant intermixing with other RP phases, including $Sr_2RuO_4$, $Sr_4Ru_3O_{10}$, and $SrRuO_3$. Based on these findings, we provide an alternative interpretation of the magnetic transitions, attributing the ferromagnetism at 170 K to $SrRuO_3$, the ferromagnetism at 100 K to $Sr_4Ru_3O_{10}$, and the unique magnetoresistance to the combined effects of the various RP phases.

**Experimental Methods**

*Thin film growth*

$Sr_3Ru_2O_7$ films were grown on a $(LaAlO_3)_{0.3}(Sr_2AlTaO_6)_{0.7}$ (LSAT) (001) (CrysTec) substrate using the hybrid MBE (Scienta Omicron) method.[25] The substrate was sequentially cleaned in acetone, methanol, and iso-propanol for 5 minutes each. The substrate was then heated to 200°C for 2 hours in the load lock chamber to evaporate any remaining solvent residue before being transferred to the growth chamber. Prior to growth, it was heated to 980°C for 20 minutes under an oxygen plasma operated at 300 W in $8\times10^{-6}$ Torr oxygen pressure to remove any adsorbed hydrocarbon-based contamination. Film was grown by co-depositing Sr metal (Sigma-Aldrich, 99.99% pure) and ruthenium acetylacetonate (American Elements, 99.99% pure) in the presence of oxygen plasma (Mantis). Sr and ruthenium acetylacetonate were supplied using effusion cells (MBE Komponenten) operated at 449°C and 154°C, respectively. Oxygen plasma pressure was $8\times10^{-6}$ Torr, and the plasma power was 300 W. The substrate temperature was maintained at 980°C, measured using a thermocouple, and the film was grown for 2 hours. The film was cooled down in the presence of oxygen plasma at a pressure of 8×10$^{-6}$ Torr until the substrate temperature reached 200˚C, to minimize the loss of oxygen from both the film and the substrate.

*Structural characterization*

In situ RHEED (STAIB Instruments) was used to monitor film surface during growth. The images were taken along the [100] azimuth of the substrate. HRXRD scan and reciprocal space map were obtained using Rigaku SmartLab XE (Rigaku Inc.) employing copper K$_α$ wavelength of 1.5406 Å. Cross-sectional specimens for the STEM analysis were fabricated using a dual-beam focused ion beam system (FIB, FEI Helios NanoLab G4). A 50 nm amorphous carbon coating was first applied to the surface of the sample through thermal evaporation. Subsequently, a 2 μm protective amorphous carbon layer was deposited over the area of interest using a Ga ion beam during the FIB milling process. The FIB was operated at an accelerating voltage of 30 kV. The STEM imaging was performed on an aberration-corrected FEI Titan G2 60–300 (S)TEM, which is equipped with a CEOS DCOR probe corrector, monochromator, and a Super-X energy dispersive X-ray spectrometer. The microscope was operated at 200 kV for image acquisition. All HAADF-STEM images were acquired using a probe current of 90 pA and convergence angle of 25.5 mrad. The inner and outer angles for the HAADF detector were 55 and 200 mrad, correspondingly.

*Electrical characterization*

Temperature-dependent longitudinal resistance, Hall resistance, and magnetoresistance were measured using a physical property measurement system (PPMS Dynacool, Quantum Design) in the Van der Pauw geometry. Aluminum wire bonded contacts were used for measurements. Resistivity was calculated from the resistance data using the Van der Pauw method and assuming a uniform thickness of 60 nm (measured from cross-sectional TEM). Magnetoresistance was measured in-plane along the [100] and [010] crystalline directions while the magnetic field is applied out-of-plane along the [001] direction. The reported magnetoresistance is the average of the [100] and [010] directions since little to no anisotropy was observed in the two in-plane directions. Hall resistance was measured in-plane for an applied B-field out-of-plane. The anomalous Hall component was obtained by subtracting the linear Hall slope obtained from high-field Hall data.

*<u>**Magnetic characterization**</u>*

Temperature- and field-dependent magnetization was measured using a superconducting quantum interference device (SQUID, Quantum Design). Temperature-dependent magnetization data was obtained by cooling the film from 300 K to 2 K in 5 T field applied along the [001] direction and measuring the magnetization during warm up using a sensing field of 100 Oe.

**Acknowledgments**

This work was supported primarily by the UMN Materials Research Science and Engineering Center (MRSEC) program under Award No. DMR-2011401. We also acknowledge partial support from the Air Force Office of Scientific Research (AFOSR) through Grant Nos. FA9550-21-1-0025. Film growth was performed using instrumentation funded by AFOSR DURIP awards FA9550-18-1-0294 and FA9550-23-1-0085. Parts of this work were carried out at the Characterization Facility, University of Minnesota, which receives partial support from the NSF through the MRSEC program under award DMR-2011401. Device fabrication was carried out at the Minnesota Nano Center, which is supported by the NSF through the National Nano Coordinated Infrastructure under award ECCS-2025124. Transport measurement at UW was supported as part of Programmable Quantum Materials, an Energy Frontier Research Center funded by the U.S. DOE, Office of Science, BES, under award DE-SC0019443.

**Author contributions:** R.C., A.K.R. and B.J. conceived the idea and designed the experiments. R.C. grew the films and characterized structurally. R.C. and A.K.R. performed electrical and magnetic characterizations. S. G. performed STEM imaging under the supervision of K.A.M. Q. J. performed in-plane angle-dependent transport measurements under the guidance of J.H.C. R.C. A.K.R. and B.J. wrote the manuscript. All authors contributed to the discussion and manuscript preparation. B.J. directed the overall aspects of the project.

**Conflict of interest:** The authors declare no conflict of interest.

**Availability of data and material:** All relevant data are included in the main text and the supplementary information. Any additional information can be requested from the corresponding authors.

**Code availability:** Not applicable.

**Supplementary information:** Supplementary material is available at DOI.

# Figures

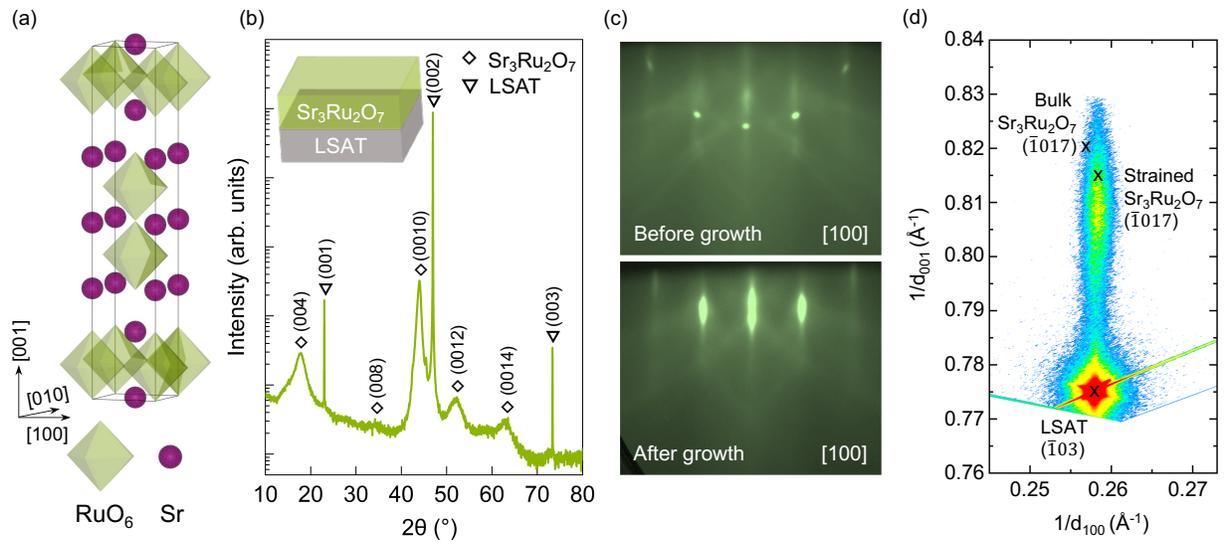

**Fig. 1.** (a) Schematic illustration of the crystal structure of $Sr_3Ru_2O_7$. lattice parameters are a = b = 3.89 Å and c = 20.7 Å. (b) HRXRD characterization of the representative $Sr_3Ru_2O_7$ film grown on LSAT (001) substrate. (c) RHEED images of the film before and after growth along the [100] azimuth of the substrate. (d) Reciprocal space map of the film around ($\bar{1}03$) substrate peak and ($\bar{1}017$) film peak. Cross marks represent the positions where peaks from the substrate and a fully strained or a fully relaxed (bulk) film is expected.

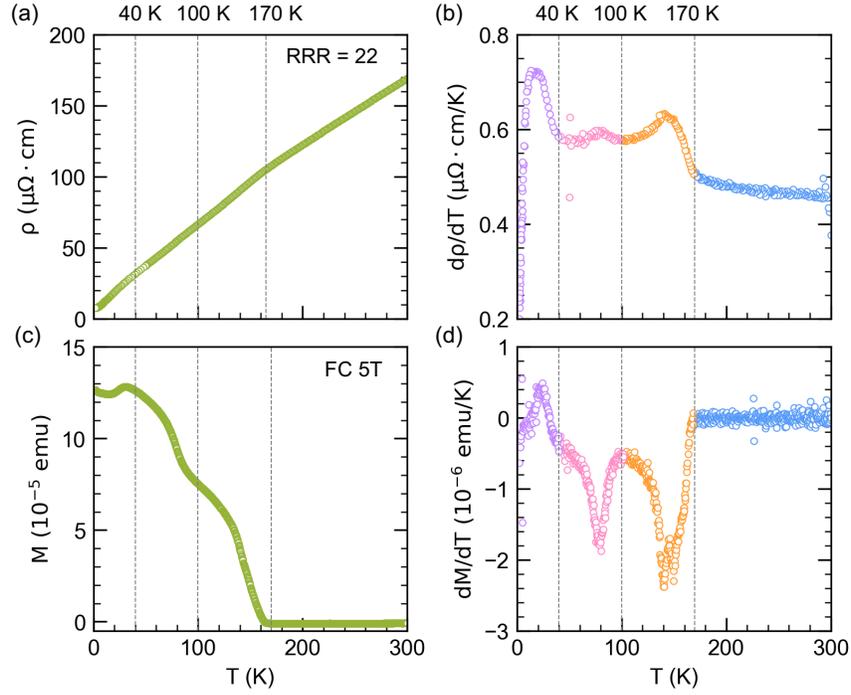

**Fig. 2.** (a) In-plane electrical resistivity of representative Sr$_3$Ru$_2$O$_7$ film from 300 K to 2 K. Residual resistivity ration of the film of 22. Vertical dashed grey lines show kink in resistivity. The exact temperature of each kink is taken from the initiation of the slope shift in the curve. (b) Derivative of the electrical resistivity with respect to temperature, to visualize slope changes in the resistivity curve. Different colors are used to visually differentiate the transitions associated with the slope change. The temperature at the onset of each peak is taken to be the transition temperature. (c) Temperature-dependent out-of-plane magnetization of the film. The films were cooled in 5 T applied field and the magnetization was measured using a 100 Oe sensing field during warming up. (d) Derivative of the magnetization with respect to temperature.

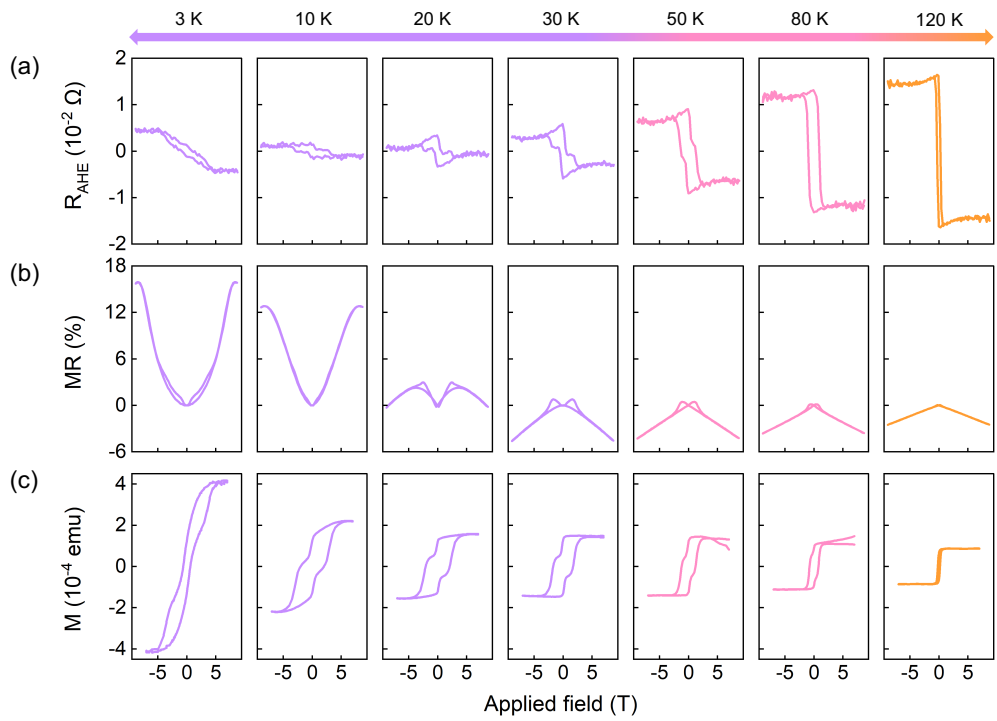

**Fig. 3.** (a) Anomalous Hall resistance (b) Magnetoresistance, and (c) Magnetization of the same representative sample at different temperatures. Applied field was out-of-plane, along the [001] direction in all of these measurements.

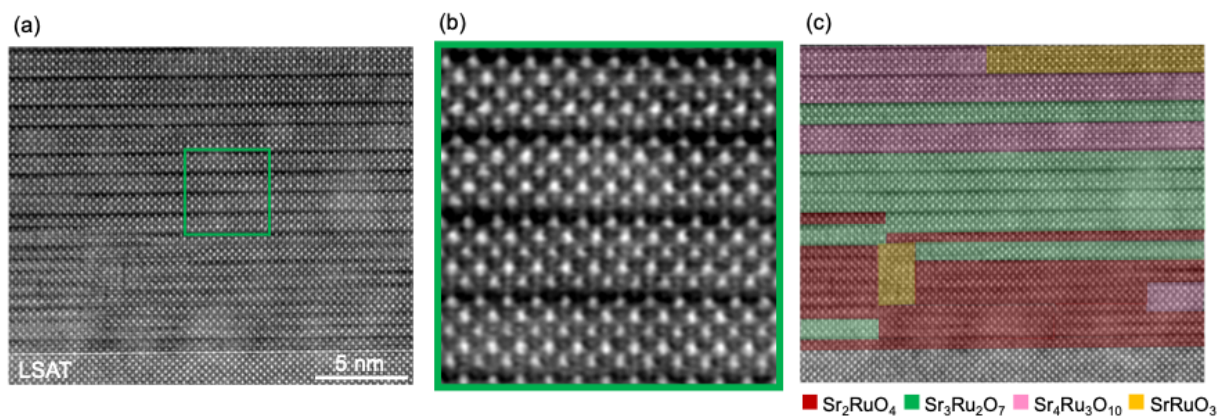

**Fig. 4.** (a) HAADF-STEM image of the representative $Sr_3Ru_2O_7$ film. The white line marks the boundary between the substrate and the film. (b) Small-area STEM image of the $Sr_3Ru_2O_7$ phase rich region, green outlined area in (a). (c) Color-coded version of (a) to show intermixing from different RP phases. The STEM images are low-pass filtered for noise reduction.